\documentclass[12pt]{article}
\input epsf
\usepackage{amssymb}
\usepackage{amsmath}
\usepackage{amsthm}
\makeatletter \@addtoreset{equation}{section} \makeatother

\def\be{\begin{equation}}
\def\ee{\end{equation}}
\def\ba{\begin{array}}
\def\ea{\end{array}}
\newtheorem*{conjecture}{Conjecture}

\newcommand{\bea}{\begin{eqnarray}}
\newcommand{\eea}{\end{eqnarray}}
\begin{document}
\hfill{}
\begin{flushright}
SU-ITP-06-20\\
SLAC-PUB-11926\\
\today
\end{flushright}
\vspace{.5cm} \vskip 1cm

\vspace{10pt}

\begin{center}
{ \Large {\bf  Exact Attractive Non-BPS STU Black Holes}} \\
\vspace{10pt}
 {\bf  Renata Kallosh$^{1}$ , Navin Sivanandam$^{1,2}$,
and Masoud Soroush$^{1,2}$
 } \\[7mm]

{\small $^1$\ Department of Physics, Stanford University, Stanford
 CA 94305-4060, USA\\
\vspace{6pt}
$^2$\ SLAC, Stanford University, Stanford CA 94309, USA}\\
\vspace{6pt} {\it kallosh@stanford.edu}\\
{\it navins@stanford.edu}\\
{\it soroush@stanford.edu} \vspace{10pt} \vspace{24pt}\\
\vspace{10pt} \abstract{We develop some properties of the non-BPS
attractive STU black hole. Our principle result is the construction
of exact solutions for the moduli, the metric and the vectors in
terms of appropriate harmonic functions. In addition, we find a
spherically-symmetric attractor carrying $p^0$ ($D6$ brane) and
$q_a$ ($D2$ brane) charges by solving the non-BPS attractor equation
(which we present in a particularly compact form) and by minimizing
an effective black hole potential. Finally, we make an argument for
the existence of multi-center attractors and conjecture that if such
solutions exist they may provide a resolution to the existence of
apparently unstable non-BPS ``attractors.''}
\end{center}
\vfill
\newpage

\tableofcontents
\newpage
\section{Introduction}
The black hole attractor story (which began with BPS states --
\cite{Ferrara:1995ih}, \cite{Strominger:1996kf},
\cite{Ferrara:1996dd}) has experienced a recent flurry of activity
with regards to the existence of non-BPS attractors
(\cite{Goldstein:2005hq}, \cite{Tripathy:2005qp}). Other interesting
recent developments include the development of the attractor
mechanism for flux vacua (\cite{Kallosh:2005ax},
\cite{Giryavets:2005nf}), the features of $\mathcal{N}=8$ attractors
(\cite{Ferrara:2006em}, \cite{Ferrara:2006yb}) and several
additional properties of both BPS and non-BPS attractors
(\cite{Bellucci:2006xz}, \cite{Kallosh:2006bx},
\cite{Kallosh:2006zs}, \cite{Bellucci:2006ew}, \cite{Sahoo:2006rp}
and \cite{Green:2006nv}).

In particular, it seems that many of the interesting properties and
features of attractive BPS configurations are shared by non-BPS
ones, so long as the black hole in question remains extremal. For
example the non-BPS attractor equation was proposed in
\cite{Kallosh:2005ax} and developed in \cite{Kallosh:2006bt}.

Such recent developments lead one to speculate on what other
attractive features may apply to non-BPS extremal black holes.
However, as we try to explore this situation, an immediate problem
arises. The first order BPS equations afford considerable
simplifications that are not present for the more general non-BPS
situation. As a result it is considerably more difficult to make
progress understanding the properties of non-BPS attractors. One way
to sidestep this calculational intractability is to consider a
simple sub-class: The STU black holes. As outlined in more detail
below, STU black holes can be constructed from compactifications of
type IIA string theory whose moduli space is described by four
homogeneous (or three inhomogeneous) co-ordinates. In particular,
our goal in this paper is to demonstrate that, for the STU black
hole, we can find exact expressions for the moduli fields, the
metric and the vectors throughout the spacetime.

It is well established that, along with giving the attractive values
of moduli, the BPS attractor equation can also be used to give an
exact expression for the moduli fields everywhere, by taking the
attractive values and replacing charges with appropriate harmonic
functions (details can be found in \cite{Behrndt:1997ny} ,
\cite{Sabra:1997dh}, \cite{Sabra:1997kq} and
\cite{Breitenlohner:1988}). While proving the equivalent statement
for non-BPS black holes is difficult, it can checked explicitly in
the STU case, where we will discover that the exact expressions for
the moduli and the metric are given by a similar prescription as for
the BPS attractors.

This paper is organized as follows. In section 2, we review the
non-BPS attractor equation for STU black holes and find a new
non-SUSY solution for a black hole carrying charges
$\Gamma=(0,q_a,p^0,0)$ (corresponding to wrapped $D2$ and $D6$
branes, and manifestly dual to the already known $D0$-$D4$ system).
Next, in section 3, we perform the calculation advertised above, and
confirm that the non-BPS attractor has its moduli described by
appropriate combinations of harmonic functions. Finally, we
speculate on the existence of multi-centered non-BPS black holes. We
argue that not only are such solutions plausible, but also that they
will exhibit a similar split attractor flow to the analogous BPS
situation and also allow us to resolve apparently unstable
``attractors'' (for example, non-BPS STU extremal black holes with
charges $\Gamma=(q_0,q_a,p^0,p^a)$, which have fixed values of the
moduli at the horizon, but are unstable to small perturbations).

\section{STU Black Holes and the Attractor Equation}

We present here a brief review of the STU  black holes and attractor
behavior -- further details can be found in \cite{Goldstein:2005hq},
\cite{Ferrara:2006yb}, \cite{Kallosh:2006bt} and
\cite{Behrndt:1996hu}.

When type IIA string theory is compactified on a $T^6$ (or,
equivalently, when M-theory is compactified on $T^7$)  one recovers
${\mathcal{N}}=8$ supergravity in d=4 with 28 vectors and 70 scalars
in the coset space $\frac{G}{H}=\frac{E_{7(7)}}{SU(8)}$, with an
on-shell U-duality symmetry $E_{7(7)}$. The STU model is a
consistent ${\mathcal{N}}=2$ truncation of a ${\mathcal{N}}=8$
model. The model depends on three complex moduli (parameterizing the
coset space $\left({SL(2, \mathbb{R})}{U(1)}\right)^3$) and has 4
vectors (one as an ${\mathcal{N}}=2$ partner of the graviton and the
others in vector multiplets).

The most general one-center BPS black hole solution of an
${\mathcal{N}}=8$ model preserving $1/8$ of the available
supersymmetry is a solution of the STU model, modulo U-duality
\cite{Andrianopoli:1997wi}.  To generate the most general solution
one has to act with $H=SU(8)$ and transform from the $4+4$ (electric
and magnetic) charges of the STU model to the $28+28$ charges of the
${\mathcal{N}}=8$ model. This corresponds to the decomposition of
the $\mathbf{56}$ of $E_{7(7)}$ with respect to $SO(4,4)\times \left
(SL(2, \mathbb{Z})\right)^3$:
\begin{equation}
\mathbf{56}\rightarrow (\mathbf{8}_v, \mathbf{2}, \mathbf{1},
\mathbf{1})\oplus (\mathbf{8}_s, \mathbf{1}, \mathbf{2},
\mathbf{1})\oplus (\mathbf{8}_{s'}, \mathbf{1}, \mathbf{1},
\mathbf{2})\oplus (\mathbf{1}, \mathbf{2}, \mathbf{2}, \mathbf{2}) \
.
\label{decomp}
\end{equation}
The charges of the STU model correspond to a singlet in $SO(4,4)$
and transform as a $(\mathbf{2}, \mathbf{2}, \mathbf{2})$
irreducible representation of $(SL(2,\mathbb{Z}))^3$ -- shown in the
last term of (\ref{decomp}). Acting with $G= E_{7(7)}$ one can
generate the generic values of scalars at infinity.

The bosonic part of the action of the four dimensional
${\mathcal{N}}=2$ supergravity  (without hypermultiplets) (in units
with $G_{N}=1$) is described by
\begin{eqnarray}\label{action}
S=\frac{1}{16\pi}\int
d^{4}x\sqrt{|g|}\Big(-\frac{{\mathcal{R}}}{2}+G_{a\bar{b}}\partial
z^{a}\cdot\partial\bar{z}^{\bar{b}}+
\mbox{Im}({\mathcal{N}}_{\Lambda\Sigma}){\mathcal{F}}^{\Lambda}\cdot{\mathcal{F}}^{\Sigma}+
\mbox{Re}({\mathcal{N}}_{\Lambda\Sigma}){\mathcal{F}}^{\Lambda}\cdot(*{\mathcal{F}})^{\Sigma}\Big)\
.
\end{eqnarray}
 The fermionic part (the gravitino and
the chiral gaugino) can be constructed via the supersymmetry
transformations:
\begin{eqnarray}\label{susytransform}
&&\delta\psi_{A\mu}=D_{\mu}\epsilon_{A}+\varepsilon_{AB}T^{-}_{\mu\nu}\gamma^{\nu}\epsilon^{B}\
,\\
&&\delta\lambda^{aA}=i\gamma\cdot\partial
z^{a}\epsilon^{A}+\frac{i}{2}\varepsilon_{AB}{\mathcal{F}^{-a}}\cdot\gamma\epsilon^{B}\
.
\end{eqnarray}
$\epsilon_{A}$ is the fermionic parameter of the transformation and
$\varepsilon_{AB}$ is the $SO(2)$ Ricci tensor. In above
transformation laws, $T^{-}$ and ${\mathcal{F}^{-a}}$ are the
two-form graviphoton and vetor multiplet field strengths
respectively. The integrals of these two two-forms on some two
cycles give us the central charge of the supersymmetry algebra and
its moduli space covariant derivative:
\begin{eqnarray}\label{centralcharge}
Z=-\frac{1}{2}\int_{S^{2}} T^{-}\ ,\
\bar{D}_{\bar{a}}\bar{Z}=-\frac{1}{2}\int_{S^{2}}
{\mathcal{F}}^{-b}G_{b\bar{a}}\ .
\end{eqnarray}

The STU models correspond to the subset of the above theories with 3
vector multiplets. The moduli space of the theory is described by
the four homogenous coordinates $X^{\Lambda}$. These coordinates
combine to give the STU prepotential  \footnote{When string theory
is compactified on some spaces which break supersymmetry down to
${\mathcal{N}}=2$, e. g. heterotic theory on $K3\times T^2$, the
prepotential (\ref{b1}) has corrections. However, when considered as
a consistent truncation of ${\mathcal{N}}=8$ model, the prepotential
is given just by eq. (\ref{b1}).}:
\begin{equation}\label{b1}
F(X)=\frac{X^{1}X^{2}X^{3}}{X^{0}}\ .
\end{equation}
The K\"{a}hler potential and superpotential of the theory can be
readily constructed from (\ref{b1}). Working with the inhomogeneous
coordinates $z^{\Lambda}=\frac{X^{\Lambda}}{X^{0}}=(1,z^{a})$ ($a\in
\{1,2,3\}$) the K\"{a}hler potential is:
\begin{equation}\label{b2}
K=-\ln\left(-i(z^{1}-\bar{z}^{1})(z^{2}-\bar{z}^{2})(z^{3}-\bar{z}^{3})\right)\
.
\end{equation}
The metric and connection on the moduli space then follow
immediately:
\begin{eqnarray}\label{b3}
G_{a\bar{b}}=-\frac{\delta_{ab}}{(z^{a}-\bar{z}^{a})^{2}}\ ,\
G^{a\bar{b}}&=&-\delta^{ab}(z^{a}-\bar{z}^{a})^{2}\ ,\
\Gamma^{a}_{aa}=-\frac{2}{z^{a}-\bar{z}^{a}}\ (\mbox{no sum on}\ a)\
.
\end{eqnarray}
Let's initially assume a static, spherically-symmetric spacetime,
with a metric of the form:
\begin{equation}
ds^2=-e^{2U(r)}dt^{2}+e^{-2U(r)}dx^idx^i
\end{equation}
In this background it is straightforward to define electric and
magnetic charges (details can be found in appendix A of
\cite{Kallosh:2006bt}) which give rise to a superpotential of the
form
\begin{eqnarray}\label{b4}
W=q_{\Lambda}X^{\Lambda}-p^{\Lambda}\partial_{\Lambda}F\ ,
\end{eqnarray}
where $q_{\Lambda}$ and $p^{\Lambda}$ are electric and magnetic
charges respectively. The set of charges
$\left(q_{0},q_{a},p^{0},p^{a}\right)$ corresponds to the charges of
$\left(D0,D2,D6,D4\right)$ branes wrapped on $\left(0,2,6,4\right)$
cycles.

An effective potential for the moduli (in this
spherically-symmetric, static spacetime )is given by:
\begin{equation}\label{b7}
V_{BH}=|DZ|^{2}+|Z|^{2}=G^{a\bar{b}}(D_{a}Z)\bar{D}_{\bar{b}}\bar{Z}+Z\bar{Z}\
.
\end{equation}
$Z$ the central charge of the supersymmetry algebra and, in our
case, is given by $Z=e^{K/2}W$. If the black hole in this solution
is extremal (i.e. has zero temperature) then it will be an attractor
for the moduli -- the values of the $z^a$ will be fixed at the
horizon, independently of their values at infinity
(\cite{Goldstein:2005hq}, \cite{Kallosh:2006bt}).

The horizon values of the moduli can be obtained by minimizing the
above effective potential -- either directly or through the
attractor equation described below. Note, however, that the extremal
point of $V_{BH}$ must be a minimum, else the attractor will be
unstable to, say, perturbations away from staticity. For BPS
attractors this condition is always satisfied, for non-BPS ones,
however, we have to check the second derivatives of $V_{BH}$
explicitly.

\subsection{The Attractor Equation}
In this section, we derive an identity using the symplectic
properties of ${\mathcal{N}}=2$ supergravity. Then, by imposing the
minimization condition of the black hole potential on the identity
we construct an algebraic relationship between the charges and the
attractive values of the moduli.

If we define the covariantly holomorphic period vector
$\Pi=(L^{\Lambda} \ M_{\Lambda})$, then the symplectic structure of
${\mathcal{N}}=2$ supergravity relates the upper and lower
components of the period vector via the vector coupling in the
following way:
\begin{eqnarray}\label{period}
M_{\Lambda}={\mathcal{N}}_{\Lambda\Sigma}L^{\Sigma}\ ,\
D_{a}M_{\Lambda}=\bar{\mathcal{N}}_{\Lambda\Sigma}D_{a}L^{\Sigma}\ ,
\end{eqnarray}
where $\Pi=(L^{\Lambda} \ M_{\Lambda})= e^{\frac{K}{2}}$
$(X^{\Lambda} \ F_{\Lambda})$,  and $(X^{\Lambda} \ F_{\Lambda})$ is
the holomorphic period vector. Now, we form a matrix by constructing
the tensor product of the period vector covariant derivative with
its complex conjugate
\begin{eqnarray}\label{identity1}
G^{a\bar{b}}D_{a}\Pi\otimes\bar{D}_{\bar{b}}\bar{\Pi}&=&G^{a\bar{b}}
\left(
  \begin{array}{cc}
    D_{a}L^{\Lambda}\bar{D}_{\bar{b}}\bar{L}^{\Sigma} & D_{a}M_{\Lambda}\bar{D}_{\bar{b}}\bar{L}^{\Sigma} \\
    D_{a}L^{\Lambda}\bar{D}_{\bar{b}}\bar{M}_{\Sigma} & D_{a}M_{\Lambda}\bar{D}_{\bar{b}}\bar{M}_{\Sigma} \\
  \end{array}
\right)\nonumber\\
&=&G^{a\bar{b}}\left(
  \begin{array}{cc}
    D_{a}L^{\Lambda}\bar{D}_{\bar{b}}\bar{L}^{\Sigma} & \bar{\mathcal{N}}_{\Lambda\Delta}
    D_{a}L^{\Delta}\bar{D}_{\bar{b}}\bar{L}^{\Sigma} \\
    {\mathcal{N}}_{\Sigma\Delta}D_{a}L^{\Lambda}\bar{D}_{\bar{b}}\bar{L}^{\Delta} &
    \bar{\mathcal{N}}_{\Lambda\Delta}{\mathcal{N}}_{\Sigma\Gamma}D_{a}L^{\Delta}\bar{D}_{\bar{b}}\bar{L}^{\Gamma} \\
  \end{array}
\right)\ ,
\end{eqnarray}
where in the second line we have used (\ref{period}). If we use the
following special geometry identity\footnote{\ This identity can be
proved by considering the inner product $\langle
D_{a}\Pi,\bar{D}_{\bar{b}}\bar{\Pi}\rangle=-iG_{a\bar{b}}$ and using
(\ref{period}).}
\begin{eqnarray}\label{identity2}
G^{a\bar{b}}D_{a}L^{\Lambda}\bar{D}_{\bar{b}}\bar{L}^{\Sigma}=-\frac{1}{2}\mbox{Im}({\mathcal{N}}^{-1})
^{\Lambda\Sigma}-\bar{L}^{\Lambda}L^{\Sigma}\ ,
\end{eqnarray}
then we find another useful identity
\begin{eqnarray}\label{identity3}
&&G^{a\bar{b}}D_{a}\Pi\otimes\bar{D}_{\bar{b}}\bar{\Pi}=
-\frac{i}{2} \left(
  \begin{array}{cc}
    0 & \mathbf{1} \\
    -\mathbf{1} & 0 \\
  \end{array}
\right)-\frac{1}{2}{\mathcal{M}}-\bar{\Pi}\otimes\Pi\ ,
\end{eqnarray}
where matrix ${\mathcal{M}}$ is defined as
\begin{eqnarray}\label{calM}
{\mathcal{M}}\equiv\left(
  \begin{array}{cc}
    \mbox{Im}({\mathcal{N}}^{-1})^{\Lambda\Sigma} & \mbox{Re}({\mathcal{N}})_{\Lambda\Delta}
    \mbox{Im}({\mathcal{N}}^{-1})^{\Delta\Sigma} \\
    \mbox{Im}({\mathcal{N}}^{-1})^{\Lambda\Delta}\mbox{Re}({\mathcal{N}})_{\Delta\Sigma} &
    \mbox{Im}({\mathcal{N}})_{\Lambda\Sigma}+\mbox{Re}({\mathcal{N}})_{\Lambda\Delta}
    \mbox{Im}({\mathcal{N}}^{-1})^{\Delta\Gamma}\mbox{Re}({\mathcal{N}})_{\Gamma\Sigma} \\
  \end{array}
\right)\ .
\end{eqnarray}
In fact, (\ref{identity3}) expresses the tensor product of the
covariant derivative of the period vector with its complex conjugate
in terms of the tensor product of the period vector itself with its
conjugate via the vector couplings.
\par
Now, assume that $\Gamma$ is the set of magnetic and electric
charges $\Gamma=(p^{\Lambda}\ q_{\Lambda})$. We define
$\tilde{\Gamma}$ by a symplectic rotation:
\begin{eqnarray}\label{gammatilde}
\tilde{\Gamma}=\left(
                 \begin{array}{cc}
                   0 & \mathbf{1} \\
                   -\mathbf{1} & 0 \\
                 \end{array}
               \right)\Gamma\ ,\
               \left(
                 \begin{array}{cc}
                   0 & \mathbf{1} \\
                   -\mathbf{1} & 0 \\
                 \end{array}
               \right)\in Sp\ (2(h^{1,1}+1),\mathbb{Z})\ .
\end{eqnarray}
Recalling (\ref{b4}), it can easily be seen that the central charge
can be expressed as $Z=\tilde{\Gamma}^{t}\cdot\Pi$. Using this form,
we can compute $G^{a\bar{b}}(D_{a}Z)\bar{D}_{\bar{b}}\bar{\Pi}$
using (\ref{identity3}):
\begin{eqnarray}\label{identity4}
G^{a\bar{b}}(D_{a}Z)\bar{D}_{\bar{b}}\bar{\Pi}=\tilde{\Gamma}^{t}\cdot
\Big(G^{a\bar{b}}D_{a}\Pi\otimes\bar{D}_{\bar{b}}\bar{\Pi}\Big)\ .
\end{eqnarray}
Thus, we have the following expression:
\begin{eqnarray}\label{identity5}
2i\bar{Z}\Pi+2iG^{a\bar{b}}(D_{a}Z)\bar{D}_{\bar{b}}\bar{\Pi}=\Gamma-i\tilde{\Gamma}^{t}\cdot{\mathcal{M}}\
.
\end{eqnarray}
It has been shown in \cite{Kallosh:2006bt}, \cite{Ferrara:1997tw},
and \cite{Ceresole:1995ca}, that the symplectic invariant $I_{1}$ is
given by:
\begin{eqnarray}\label{identity6}
I_{1}=|Z|^{2}+|DZ|^{2}=-\frac{1}{2}\
\tilde{\Gamma}^{t}\cdot{\mathcal{M}}\cdot\tilde{\Gamma}\ .
\end{eqnarray}
Finally, we differentiate the above expression with respect to the
charges and substitute for (\ref{identity5}):
\begin{equation}\label{ee1}
\Gamma+i\frac{\partial
I_{1}}{\partial\tilde{\Gamma}}=2i\bar{Z}\Pi+2iG^{a\bar{b}}(D_{a}Z)\bar{D}_{\bar{b}}\bar{\Pi}\
.
\end{equation}

So far, everything we have said is generically true and independent
of the detailed model. However, if we restrict ourselves to STU
models, then the symplectic invariant $I_{1}$ is a function of
charges. It is given by $I_{1}=\sqrt{|{\mathcal{W}}(\Gamma)|}$, in
which ${\mathcal{W}}$ is given by:
\begin{equation}\label{hyperdet}
{\mathcal{W}}(\Gamma)=4((p^{1}q_{1})(p^{2}q_{2})+
(p^{1}q_{1})(p^{3}q_{3})+(p^{2}q_{2})(p^{3}q_{3}))
-(p^{\Lambda}q_{\Lambda})^{2}-4p^{0}q_{1}q_{2}q_{3}+4q_{0}p^{1}p^{2}p^{3}\
.
\end{equation}
In \cite{Behrndt:1997ny}, Berhrndt et al. demonstrated that
$\mathcal{W}(\Gamma)$ is an $[SL(2,\mathbb{Z})]^3$ invariant which
uniquely determines the form of the STU metric and further
$-\mathcal{W}$ has been established as the Cayley hyperdeterminant
by Duff in \cite{Duff:2006uz}. The fact we must use the absolute
value of $\mathcal{W}$ in order that this expression be valid for
non-BPS attractors was discussed in \cite{Kallosh:2006zs} -- here it
was observed that $\mathcal{W}(\Gamma)$ is positive for BPS
attractors and negative for non-BPS ones.

Recalling that at an attractor point the effective black hole
potential will be minimized (see \cite{Ferrara:1997tw},
\cite{Goldstein:2005hq}, \cite{Kallosh:2006bt}) we note that
extremization of (\ref{b7}) gives:
\begin{equation}\label{ee2}
2\bar{Z}D_{\hat{a}}Z+i C_{\hat{a}\hat{b}\hat{c}}\eta^{\hat{b}\hat{\bar{d}}}\eta^{\hat{c}\hat{\bar{e}}}
(\bar{D}_{\hat{\bar{d}}}\bar{Z})(\bar{D}_{\hat{\bar{e}}}\bar{Z})=0\
.
\end{equation}
Hatted indices are for the tangent space and
$\eta^{\hat{a}\hat{\bar{b}}}$ is a flat Euclidean metric.
 Here $G^{a \bar a}= e^a_{\hat a} \eta^{\hat a \hat {\bar a}} e_{\hat {\bar a}}^{\bar a}$.
The rank 3 completely symmetric tensor
$C_{\hat{a}\hat{b}\hat{c}}$ in the tangent space is related to the covariantly
holomorphic symmetric  curved space symmetric tensor  $C _{abc}$   as follows:
$C_{\hat{a}\hat{b}\hat{c}} = e^a_{\hat a}e^b_{\hat b}
 e^c_{\hat c}C _{abc}$. Here the rank 3 completely symmetric covariantly holomorphic  tensor of
 the special K\"{a}hler geometry, $C _{abc}$, is defined as follows
 \cite{Ceresole:1995ca}:
\begin{equation}
C_{abc} = e^K (\partial_a X^A \partial_b X^B \partial_c X^C)
\partial_A \partial_B\partial_C F \ .
\end{equation}
For the STU model with the prepotential (\ref{b1}) the non-vanishing
components of the symmetric 3-tensor are
$C_{\hat{1}\hat{2}\hat{3}}=1$

Note that the covariantly holomorphic symmetric 3-tensor $C _{abc}$
has a chiral weight 2 under K\"{a}hler transformations whereas
$C_{\hat{a}\hat{b}\hat{c}}$ in the tangent space has a chiral weight
0.

The chiral $U(1)$ weights of various objects living on K\"{a}hler
manifolds are defined as follows. The K\"{a}hler potential
transforms as:
\begin{equation}
K\rightarrow K+ f(z) + \bar f (\bar z) \ .
\end{equation}
The covariant derivative of an object with the chiral weight $p$ is
given by $D\Phi=(d+ip{\cal Q})\Phi$ where the K\"{a}hler connection
is given by ${\cal Q}=-{i\over 2}( \partial_a K dz^a -
\partial_{\bar a} K d\bar z^{\bar a})$. In components this means:
\begin{equation}
D_a \Phi = (\partial_a + {1\over 2} p\, \partial_a K) \Phi \ .
\end{equation}
Thus the chiral weight of $Z$ is 1, $\bar Z$ is -1, $e^a_{\hat a}$
is -2/3 and $e_{\hat {\bar a}}^{\bar a}$ is 2/3. This means, in
particular, that under K\"{a}hler transformations the central charge
$Z$ as well as its covariant derivative transform by a phase:
\begin{equation}
Z\rightarrow Z \, e^{\bar f-f\over 2}\ ,  \qquad D_{\hat a} Z
\rightarrow  D_{\hat a} Z \, e^{\bar f-f\over 6} \ .
\label{chiral}
\end{equation}
Further, observe that (\ref{ee2}) has a chiral weight -2/3 whereas
(\ref{ee1}) has a vanishing chiral weight. In \cite{Ferrara:2006yb}
(\ref{ee2}) was studied by embedding the ${\mathcal{N}}=2$
supergravity into an ${\mathcal{N}}=8$ one. Following the same
method we define $Y_{0}\equiv iZ$, and $Y_{a}\equiv
\bar{D}_{\hat{\bar{a}}}\bar{Z}$. This gives us the solution:
\begin{equation}\label{ee3}
Y_{0}=\rho e^{i(\pi-3\phi)}\ ,\ Y_{a}=\rho e^{i\phi}\ .
\end{equation}
Here the arbitrary phase $\phi$ reflects the possibility to perform
a K\"{a}hler transformations in agreement with the weights shown in
(\ref{chiral}). (\ref{ee1}) can be rewritten in the following form:
\begin{equation}\label{ee4}
\Gamma+i\frac{\partial
I_{1}}{\partial\tilde{\Gamma}}=2\rho(\Pi+i\sum_{\hat{a}=1}^{3}\bar{D}_{\hat{\bar{a}}}\bar{\Pi})\
\end{equation}
The arbitrary phase $\phi$ drops from this equation since the left
hand side of eq. (\ref{ee1} is manifestly K\"{a}hler invariant and
in the right hand side the phase cancels in each term separately.

Now, if we divide (\ref{ee4}) by its zeroth component, then we
follow \cite{Ferrara:2006yb} and obtain:
\begin{eqnarray}\label{e1}
\frac{p^{\Lambda}+i\frac{\partial I_{1}}{\partial
q_{\Lambda}}}{p^{0}+i\frac{\partial I_{1}}{\partial
q_{0}}}&=&\frac{L^{\Lambda}+i\sum_{\hat{a}=1}^{3}\bar{D}_{\hat{\bar{a}}}\bar{L}^{\Lambda}}
{L^{0}+i\sum_{\hat{a}=1}^{3}\bar{D}_{\hat{\bar{a}}}\bar{L}^{0}} \\
\frac{q_{\Lambda}-i\frac{\partial I_{1}}{\partial
p^{\Lambda}}}{q_{0}-i\frac{\partial I_{1}}{\partial
p^{0}}}&=&\frac{M_{\Lambda}+i\sum_{\hat{a}=1}^{3}\bar{D}_{\hat{\bar{a}}}\bar{M}_{\Lambda}}
{M_{0}+i\sum_{\hat{a}=1}^{3}\bar{D}_{\hat{\bar{a}}}\bar{M}_{0}}\ .
\end{eqnarray}
In fact, these equations are dual to each other and either set is
sufficient to completely determine the moduli at the horizon.

We now illustrate that an even simpler form of these equations can
be found, as the r.h.s simplifies significantly. Consider
(\ref{e1}), since (with the vielbein $e_{\hat{b}}^{a}$ given by
$e_{\hat{b}}^{a}=i\delta^{a}_{b}(z^{a}-\bar{z}^{a})$):
\begin{equation}\label{e3}
D_{\hat{b}}L^{c}=e_{\hat{b}}^{a}D_{a}L^{c}=e^{K/2}e_{\hat{b}}^{a}(\delta^{c}_{a}+(\partial_{a}K)z^{c})\
,
\end{equation}
we can readily establish:
\begin{equation}\label{e5}
L^{c}+i\sum_{\hat{a}=1}^{3}\bar{D}_{\hat{\bar{a}}}\bar{L}^{c}=-2e^{K/2}\bar{z}^{c}\
,\
L^{0}+i\sum_{\hat{a}=1}^{3}\bar{D}_{\hat{\bar{a}}}\bar{L}^{0}=-2e^{K/2}\
.
\end{equation}
As expected the above is independent of charges, and the values of
the moduli at the horizon for any $\Gamma$ are thus given by:
\begin{equation}\label{e6}
z^{\Lambda}(\Gamma)=\frac{p^{\Lambda}-i\frac{\partial
I_{1}(\Gamma)}{\partial q_{\Lambda}}}{p^{0}-i\frac{\partial
I_{1}(\Gamma)}{\partial q_{0}}}\ .
\end{equation}
This appears to be the simplest form of the non-BPS attractor
equation for the STU model. In order to obtain eq. (2.34), we have used the gauge in which $X^{0}$ was real.  However, the attractor equation can  be
generalized\footnote{ We are grateful to E. Gimon who suggested to look for a more general form of eq. (2.34) which would include (D0-D6) system.} to the   case in which
$X^{0}=e^{-\frac{3i\delta}{4}}$, with real $\delta$. In this  gauge  the attractor equation generalizing the one in eq. (2.34) becomes
\begin{eqnarray}\label{attractor-general}
\frac{z^{c}-e^{i\delta}(2\bar{z}^{c}+z^{c})}{1-3e^{i\delta}}=
\frac{p^{c}+i\frac{\partial I_{1}(\Gamma)}
{\partial q_{c}}}{p^{0}+i\frac{\partial I_{1}(\Gamma)}{\partial q_{0}}}\ ,
\label{general} \end{eqnarray}
and similar equation for component $M_{\Lambda}$. In case that  $\delta= 2n\pi$ eq. (\ref{general}) reduces to eq. (2.34). This gives a correct solution for
 (D2-D6) and (D0-D4) systems. However, if one considers (D0-D6) system, then
 eq. (\ref{general})
 requires $\delta=(2n+1)\pi$.

For the BPS case $Y_{a}$ vanishes and so a similar procedure to that
used above will result in an attractor equation of the form of
(\ref{ee4}), but with the r.h.s. containing only the first term.
Dividing such an expression by its zeroth component we would obtain:
\begin{equation}\label{ee6}
z^{\Lambda}(\Gamma)=\frac{p^{\Lambda}+i\frac{\partial
I_{1}(\Gamma)}{\partial q_{\Lambda}}}{p^{0}+i\frac{\partial
I_{1}(\Gamma)}{\partial q_{0}}}\ .
\end{equation}
Of course, this is not simply the complex conjugate of (\ref{e6}),
as in the BPS case $\mathcal{W}>0$, whilst for the non-BPS case
$\mathcal{W}<0$ (recall $I_1=\sqrt{|\mathcal{W}|}$).

\subsection{Solutions to the Attractor Equation}
It is relatively straightforward to obtain solutions to the above
attractor equation when we restrict ourselves to a system with only
$D0$ and $D4$ branes ($q_0$ and $p^a$ charges). This attractor was
found in \cite{Tripathy:2005qp} by minimizing $V_{BH}$ -- in fact,
the solution here is not restricted to STU, i.e. there may be any
number of the moduli. We consider here, though, the complementary
solution with $D6$ and $D2$ branes ($p^0$ and $q_a$ charges). The
superpotential is then given by:
\begin{eqnarray}\label{b5}
W(z^1,z^2,z^3)=q_{a}z^{a}+p^{0}z^{1}z^{2}z^{3}\ .
\end{eqnarray}
It is also straightforward to write down the symplectic invariant
$I_1$:
\begin{equation}\label{I1D2D6}
I_1=\sqrt{\pm4p^{0}q_{1}q_{2}q_{3}}\ .
\end{equation}
A positive sign under the square root corresponds to the non-BPS
attractor and a negative one to the BPS one, since, as discussed
above, $\mathcal{W}(\Gamma)$ is generally positive for charges
corresponding to BPS solutions and negative for those corresponding
to non-BPS ones (it is evident from the form of the solutions below
that $p^{0}q_{1}q_{2}q_{3}<0$ for BPS attractors
$p^{0}q_{1}q_{2}q_{3}>0$ otherwise).

>From (\ref{I1D2D6}) it is trivial to use (\ref{ee6}) to obtain the
BPS attractive moduli values as:
\begin{eqnarray}\label{bpsd2d6}
z^{1}=-i\sqrt{-\frac{q_{2}q_{3}}{p^{0}q_{1}}}\ ,\
z^{2}=-i\sqrt{-\frac{q_{1}q_{3}}{p^{0}q_{2}}}\ ,\
z^{3}=-i\sqrt{-\frac{q_{1}q_{2}}{p^{0}q_{3}}}\ .
\end{eqnarray}
The non-BPS (from (\ref{e6})) results have a similar form:
\begin{eqnarray}\label{nonbpsd2d6}
z^{1}=-i\sqrt{\frac{q_{2}q_{3}}{p^{0}q_{1}}}\ ,\
z^{2}=-i\sqrt{\frac{q_{1}q_{3}}{p^{0}q_{2}}}\ ,\
z^{3}=-i\sqrt{\frac{q_{1}q_{2}}{p^{0}q_{3}}}\ .
\end{eqnarray}
These results have been confirmed by minimizing $V_{BH}$ and by
using the form of the attractor equation given in
\cite{Kallosh:2006bt} -- these details of those calculations can be
found in appendices A.1 and A.2. It's worth noting at this juncture
that care must be taken with these solutions to ensure that their
signs are such that $e^K$ is positive. This point is discussed in
somewhat more detail in the appendix. We also observe that
(\ref{e6}) and (\ref{ee6}) can be used to find the attractor values
of the moduli for the $D0$-$D4$ system addressed in
\cite{Tripathy:2005qp} and \cite{Kallosh:2006bt}. As expected the
BPS answers are:
\begin{eqnarray}\label{bpsq0pi}
z^{1}=-i\sqrt{\frac{q_{0}p^{1}}{p^{2}p^{3}}}\ ,\
z^{2}=-i\sqrt{\frac{q_{0}p^{2}}{p^{3}p^{1}}}\ ,\
z^{3}=-i\sqrt{\frac{q_{0}p^{3}}{p^{1}p^{2}}}\ .
\end{eqnarray}
And the non-BPS:
\begin{eqnarray}\label{nonbpsq0pi}
z^{1}=-i\sqrt{-\frac{q_{0}p^{1}}{p^{2}p^{3}}}\ ,\
z^{2}=-i\sqrt{-\frac{q_{0}p^{2}}{p^{3}p^{1}}}\ ,\
z^{3}=-i\sqrt{-\frac{q_{0}p^{3}}{p^{1}p^{2}}}\ .
\end{eqnarray}

As mentioned above, for a non-BPS attractor we are required to
confirm that the extremal point of the potential does correspond to
a minima. It turns out that this is a somewhat subtle and involved
calculation. The details can be found in appendix A.3, but the
short-answer to the question of stability is (as in the $D0$-$D4$
case discussed in \cite{Tripathy:2005qp}): ``It's stable.''

Before moving on to a discussion of the complete solutions for the
moduli fields we summarize our knowledge of STU black hole
attractors:
\begin{itemize}
\item All extremal STU black holes can exhibit attractor behavior,
with the values of the moduli at the horizon found by extremizing
the effective potential $V_{BH}$.
\item Only those systems with extremum of the potential a minimum
will form stable attractors. These systems include those with
$\Gamma=(q_0,0,0,p^a)$ and those with $\Gamma=(0,q_a,p^0,0)$.
Unstable ``attractors'' include those with all four types of charge,
$\Gamma=(q_0,q_a,p^0,p^a)$; this result was established in
\cite{Tripathy:2005qp} (see appendix A.3 for details).
\end{itemize}
We shall return to the issue of stability when we consider
multi-centered non-BPS black holes in section 4. Now, however, we
move on to the general solution for the moduli.

\section{The Exact Non-BPS Solution}
We now demonstrate explicitly the construction of exact solutions
for the fields in the non-BPS attractor. We begin with a general
discussion, and then write down and prove the form of the solutions.
\subsection{Harmonic Functions and U-Duality}
So far, we have calculated the value of moduli at the horizon of the
supersymmetric and non-supersymmetric STU black holes via the
attractor mechanism. It turns out, however, that we can do better
than that -- we can find solutions to the equations of motion for
the moduli, allowing us to obtain their values everywhere. This has
already been done for the BPS case (details can be found in
\cite{Behrndt:1997ny}, \cite{Sabra:1997dh}, \cite{Sabra:1997kq} and
\cite{Bates:2003vx}), and we will proceed in an analogous fashion.

We work with the single-center spherically-symmetric, static metric
ansatz:
\begin{equation}
ds^2=-e^{2U}dt^2+e^{-2U}dx^idx^i\ .
\end{equation}
The basic idea is that one takes the horizon ($r=r_h$) values of the
moduli, $z^a(q_{\Lambda},p^{\Lambda})$ and replaces the charges with
harmonic functions:
\begin{equation}\label{harm1}
z^{\Lambda}(H(\mathbf{x}))=\frac{H^{\Lambda}-i\frac{\partial
I_{1}(H)}{\partial H_{\Lambda}}}{H^{0}-i\frac{\partial
I_{1}(H)}{\partial H_{0}}}\ .
\end{equation}
The metric is found through:
\begin{equation}
e^{-2U}=I_1(\mathbf{H})
\end{equation}
The harmonic functions $\mathbf{H}$ are:
\begin{equation}\label{f4}
\mathbf{H}(\tau)=\left(H^{\Lambda},H_{\Lambda}\right)=
\left(\Gamma^{\Lambda},\Gamma_{\Lambda}\right)\tau+\left(h^{\Lambda},h_{\Lambda}\right)\
,\ \tau=\frac{1}{|r-r_h|}\ .
\end{equation}
$I_1$ is the symplectic invariant defined above and there is a
constraint on $\mathbf{h}$:
$\left<\mathbf{h},\mathbf{\Gamma}\right>=0$. $\mathbf{H}$ arises
from the solution to the vector equations of motion where the vector
fields can be represented by the doublet $(F^\Lambda, G_\Lambda)$.
These fields are not independent:
\begin{equation}\label{relation}
G_{\Lambda} = \mathrm{Re}\,\mathcal{N}_{\Lambda\Sigma}(z, \bar z)
F^{\Lambda}-\mathrm{Im}\,\mathcal{N}_{\Lambda\Sigma}(z,\bar{z})*F^\Lambda
\end{equation}
The equations of motion then require a vector field potential given
by $(\mathcal{A}^{\Lambda},\mathcal{B}_{\Lambda})$ where
$F^{\Lambda}= d\mathcal{A}^{\Lambda}$ and
$G_{\Lambda}=d\mathcal{B}_{\Lambda}$. The vector fields are related
to the harmonic functions by:
\begin{equation}
F^{\Lambda}_{mn}=\frac{1}{2}\epsilon_{mnp}\partial_pH^{\Lambda}\ ,\
G_{\Lambda mn}=\frac{1}{2}\epsilon_{mnp}\partial_pH_{\Lambda}\ .
\end{equation}
The timelike components of $F$ and $G$ are fixed in terms of the
spatial ones through (\ref{relation}).

That this result holds in the BPS case has been established
explicitly; there is, however, no such proof for non-BPS black
holes. Instead, we shall apply the above algorithm to generate an
ansatz which can be checked explicitly using the STU equations of
motion. Before we do so, though, we offer an argument as to why one
might expect the above prescription to work.

There exists a manifest $[SL(2,\mathbb{Z})]^3$ symmetry of the
equations of motion which mixes the Maxwell equations and Bianchi
identities:
\begin{equation}\label{vect}
\begin{array}{c}
\left(
\begin{array}{c}
{\cal A}^\Lambda(\vec x)\\
{\cal B}_\Lambda (\vec x)
\end{array}
\right)'
\end{array}= \begin{array}{cc}
\left(
\begin{array}{cc}
A& B\\
C& D
\end{array}
\right)
\end{array}\begin{array}{c}
\left(
\begin{array}{c}
{\cal A}^\Lambda(\vec x)\\
{\cal B}_\Lambda(\vec x)
\end{array}
\right)
\end{array}
\end{equation}
Here $\begin{array}{cc} \left(
\begin{array}{cc}
A& B\\
C& D
\end{array}
\right)
\end{array}$ is a global $Sp(8, \mathbb{Z})$ matrix
such that  $A^T C- C^T A= B^T D- D^T B=0$ and $A^T D- C^T B=1$. Only
a subgroup of this symmetry, corresponding to $[SL(2,\mathbb{Z})]^3$
duality symmetry, is unbroken at the level of solutions to the
equations of motion. The details of the embedding of
$[SL(2,\mathbb{Z})]^3$ into $Sp(8, \mathbb{Z})$ are given in
\cite{Behrndt:1996hu} and \cite{Duff:1995sm}. Simultaneously with
the duality transformations of the vector potential doublet the 3
moduli also undergo a standard fractional transformation of the
type:
\begin{equation}\label{duality}
(z^i(\vec{x}))' = \frac{a\, z^i (\vec{x})+b}{c\, z^i(\vec{x})+d}\ .
\end{equation}
This transformation of the moduli follows from the consistency of
equation (\ref{relation}) and leaves the metric invariant
\begin{equation}
(g_{\mu\nu}(\vec{x}))'= g_{\mu\nu}(\vec{x})\ ,
\end{equation}
allowing us to generate from one solution all the others related
through duality. Clearly, such a symmetry places strong constraints
on the form of the moduli fields. Further, our explorations of
functional forms have been unable to find any solutions with the
appropriate symmetry other than those given by (\ref{harm1}), when
we simultaneously impose the obvious criteria that the moduli give
the correct horizon values, remain bounded everywhere and have
derivatives that do the same while also vanishing at the horizon
(see \cite{Kallosh:2006bt} for the origins of these restrictions).

\subsection{The Proof}
We shall begin by establishing that the appropriate ansatz solves
the $D2$-$D6$ brane system and then find the general (almost -- see
below for details) solution through a symmetry argument.

\subsubsection{The $D2$-$D6$ System}
The four dimensional stationary, spherically-symmetric effective
Lagrangian of the Maxwell-Einstein action (derived from
(\ref{action})) is:
\begin{eqnarray}\label{f1}
{\mathcal{L}}(U(\tau),z^{a}(\tau),\bar{z}^{\bar{a}}(\tau))=
\left(\dot{U}^{2}+G_{a\bar{b}}\dot{z}^{a}\dot{\bar{z}}^{\bar{b}}+e^{2U}V_{BH}\right)\
.
\end{eqnarray}
$\tau$ is the inverse of the radial coordinate (such that the
horizon is at $\tau=-\infty$ and the derivatives denote
differentiation with respect to $\tau$. The gravitational and scalar
equations of motion derived from the above Lagrangian (\ref{f1}) are
\begin{eqnarray}\label{f2}
&&\ddot{U}=e^{2U}V_{BH}\ ,\\
&&\ddot{z}^{a}+\Gamma^{a}_{bc}\dot{z}^{b}\dot{z}^{c}=e^{2U}G^{a\bar{b}}\partial_{\bar{b}}V_{BH}\
.
\end{eqnarray}
In addition, there is also a constraint on the system:
\begin{eqnarray}\label{f3}
\dot{U}^{2}+G_{a\bar{b}}\dot{z}^{a}\dot{\bar{z}}^{\bar{b}}-e^{2U}V_{BH}=c^{2}\
,
\end{eqnarray}
where $c^{2}=2ST=0$ for extremal black holes.

As described above, to generate solutions to these equations we take
the horizon values of the moduli, (\ref{nonbpsd2d6}), and replace
the charges with appropriate harmonic functions:
\begin{eqnarray}\label{f5}
e^{-2U}&=&2\sqrt{H^{0}H_{1}H_{2}H_{3}}\nonumber\\
z^{1}&=&-i\sqrt{\frac{H_{2}H_{3}}{H^{0}H_{1}}}\nonumber\\
z^{2}&=&-i\sqrt{\frac{H_{1}H_{3}}{H^{0}H_{2}}}\nonumber\\
z^{3}&=&-i\sqrt{\frac{H_{1}H_{2}}{H^{0}H_{3}}}\ .
\end{eqnarray}
Substitution followed by explicit calculation of the terms in the
gravitational equation of motion gives:
\begin{eqnarray}\label{f6}
\ddot{U}e^{-2U}=V_{BH}&=&\frac{1}{2}\Big[\frac{q_{1}^{2}}{(H_{1})^{3/2}}\sqrt{H^{0}H_{2}H_{3}}
+\frac{q_{2}^{2}}{(H_{2})^{3/2}}\sqrt{H^{0}H_{1}H_{3}}\nonumber\\
&&+\frac{q_{3}^{2}}{(H_{3})^{3/2}}\sqrt{H^{0}H_{1}H_{2}}+
\frac{(p^{0})^{2}}{(H^{0})^{3/2}}\sqrt{H_{1}H_{2}H_{3}}\Big]\ .
\end{eqnarray}
In a somewhat more involved calculation we can also verify that the
ansatz solves the scalar equations of motion. Picking the 1
direction, we have:
\begin{eqnarray}\label{f7}
\ddot{z}^{1}+\Gamma^{1}_{11}(\dot{z}^{1})^{2}&=&e^{2U}G^{1\bar{1}}\partial_{\bar{1}}V_{BH}\nonumber\\
&=&e^{2U+K}G^{1\bar{1}}
\Big(G^{2\bar{2}}(\bar{D}_{\bar{1}}\bar{D}_{\bar{2}}\bar{W})D_{2}W+
G^{3\bar{3}}(\bar{D}_{\bar{1}}\bar{D}_{\bar{3}}\bar{W})D_{3}W+2(\bar{D}_{\bar{1}}\bar{W})W\Big)\
.
\end{eqnarray}
Once again substitution and tedium give the result we desire:
\begin{eqnarray}\label{f8}
\ddot{z}^{1}+\Gamma^{1}_{11}(\dot{z}^{1})^{2}&=&e^{2U}G^{1\bar{1}}\partial_{\bar{1}}V_{BH}\nonumber\\
&=&\frac{i}{2}\left(\frac{H_{2}H_{3}}{H^{0}H_{1}}\right)^{-3/2}\Big[q_{2}^{2}(H^{0}H_{1}H_{3})^{2}
+q_{3}^{2}(H^{0}H_{1}H_{2})^{2}\nonumber\\
&&-q_{1}^{2}(H^{0}H_{2}H_{3})^{2}-(p^{0})^{2}(H_{1}H_{2}H_{3})^{2}\Big]\
.
\end{eqnarray}
Finally, we check the constraint:
\begin{equation}\label{f9}
\dot{U}^{2}+G_{a\bar{b}}\dot{z}^{a}\dot{\bar{z}}^{\bar{b}}=e^{2U}V_{BH}=\frac{1}{4}
\left[\frac{q_{1}^{2}}{H_{1}^{2}}+\frac{q_{2}^{2}}{H_{2}^{2}}+\frac{q_{3}^{2}}{H_{3}^{2}}
+\frac{(p^{0})^{2}}{(H^{0})^{2}}\right]\ .
\end{equation}

\subsubsection{The Generic System: (D0,D2,D4,D6)}
To generalize the previous result, we need to establish that the
prescription of the previous section, i.e. the substitution of
harmonic functions for the electric and magnetic charges in the
values of moduli at the horizon, solves the equations of motion when
the moduli and metric are given by the expressions above:
\begin{equation}\label{e7}
z^{\Lambda}(H(\mathbf{x}))=\frac{H^{\Lambda}-i\frac{\partial
I_{1}(H)}{\partial H_{\Lambda}}}{H^{0}-i\frac{\partial
I_{1}(H)}{\partial H_{0}}}\ ,\
e^{-2U}=\sqrt{|\mathcal{W}(\mathbf{H})|}\ .
\end{equation}
Given the hugely increased complexity of $I_1$ when all charges are
present (not to mention $V_{BH}$) an explicit calculation would
painful. Accordingly, we're going to do something else.

The symplectic invariance of special geometry ensures that the
Lagrangian of our theory has an $Sp(8,{\mathbb{Z}})$ symmetry, which
reduces to $[SL(2,{\mathbb{Z}})]^{3}$ at the level of the equations
of motion. This symmetry group can be used to generate solutions for
generic charges by rotating the expressions for the moduli in the
$D2$-$D6$ system (eq. (\ref{f5})) obtained above. To do this we take
the following element of $[SL(2,{\mathbb{Z}})]^{3}$:
\begin{equation}\label{e77}
\left(
  \begin{array}{cc}
    a & b \\
    c & d \\
  \end{array}
\right)\otimes \left(
\begin{array}{cc}
1 & 0 \\
0 & 1 \\
\end{array}
\right)\otimes \left(
  \begin{array}{cc}
    1 & 0 \\
    0 & 1 \\
  \end{array}
\right)\ ,
\end{equation}
in which $ad-bc=1$. Under this transformation, the charges and
harmonic functions of the $D2$-$D6$ system transform in the
following way
\begin{equation}\label{e8}
\left(
  \begin{array}{c}
    \tilde{p}^{0} \\
    \tilde{p}^{1} \\
    \tilde{p}^{2} \\
    \tilde{p}^{3} \\
    \tilde{q}_{0} \\
    \tilde{q}_{1} \\
    \tilde{q}_{2} \\
    \tilde{q}_{3} \\
  \end{array}
\right)=
  \left(
    \begin{array}{c}
      dp^{0} \\
      bp^{0} \\
      cq_{3} \\
      cq_{2} \\
      -bq_{1} \\
      dq_{1} \\
      aq_{2} \\
      aq_{3} \\
    \end{array}
  \right)\ ,\
  \left(
    \begin{array}{c}
    \tilde{H}^{0} \\
    \tilde{H}^{1} \\
    \tilde{H}^{2} \\
    \tilde{H}^{3} \\
    \tilde{H}_{0} \\
    \tilde{H}_{1} \\
    \tilde{H}_{2} \\
    \tilde{H}_{3} \\
    \end{array}
  \right)=\left(
    \begin{array}{c}
      dH^{0} \\
      bH^{0} \\
      cH_{3} \\
      cH_{2} \\
      -bH_{1} \\
      dH_{1} \\
      aH_{2} \\
      aH_{3} \\
    \end{array}
  \right)\ .
\end{equation}
Further, we know that under transformation (\ref{e77}), the moduli
coordinates transform as
\begin{equation}\label{e9}
z^{1}\mapsto \tilde{z}^{1}=\frac{az^{1}+b}{cz^{1}+d}\ ,\
z^{2}\mapsto \tilde{z}^{2}=z^{2}\ ,\ z^{3}\mapsto
\tilde{z}^{3}=z^{3}\ .
\end{equation}
The solution of equations of motion for the case of generic
charges\footnote{\ We observe that the charges (and relevant
harmonic function) are not completely independent and we have these
two relations among them:
$\tilde{p}^{3}\tilde{q}_{3}=\tilde{p}^{2}\tilde{q}_{2}$ and
$\tilde{p}^{1}\tilde{q}_{1}=-\tilde{p}^{0}\tilde{q}_{0}$. If we also
make nontrivial $SL(2,{\mathbb{Z}})$ transformations on the second
and third moduli this dependence disappears. On the other hand, the
above analysis shows that this procedure can be applied for
successive transformations. Therefore, we do not lose the generality
of the argument.} can be swiftly obtained from (\ref{e9}).

To check whether (\ref{e7}) (with a set of generic charges)
coincides with (\ref{e9}) it is sufficient to establish that
(\ref{e9}) will solve the equations of motion. To do this we first
we observe, as proved in \cite{Behrndt:1997ny}, that the
hyper-determinant ${\mathcal{W}}$ is invariant under (\ref{e77}). In
our particular case this is simple to see from:
\begin{eqnarray}\label{e10}
{\mathcal{W}}(H)\mapsto
\tilde{{\mathcal{W}}}(\tilde{H})=-4(ad-bc)^{2}H^{0}H_{1}H_{2}H_{3}={\mathcal{W}}(H)\
.
\end{eqnarray}
Next, using (\ref{e7}), we obtain the modulus along the first
direction of the moduli space for the generic case as
\begin{eqnarray}\label{e11}
\tilde{z}^{1}=\frac{\tilde{H}^{1}\sqrt{|{\mathcal{W}}|}+i(\tilde{H}^{1}(\tilde{H}^{2}\tilde{H}_{2}
+\tilde{H}^{3}\tilde{H}_{3}-\tilde{H}^{0}\tilde{H}_{0}-
\tilde{H}^{1}\tilde{H}_{1})-2\tilde{H}^{0}\tilde{H}_{2}\tilde{H}_{3})}
{\tilde{H}^{0}\sqrt{|{\mathcal{W}}|}+i(2\tilde{H}^{1}\tilde{H}^{2}\tilde{H}^{3}-\tilde{H}^{0}(\tilde{H}^{\Lambda}
\tilde{H}_{\Lambda}))}\ .
\end{eqnarray}
This can be reexpressed in terms of the harmonic functions of the
$D2$-$D6$ system:
\begin{equation}\label{e12}
\tilde{z}^{1}=\frac{b\sqrt{H^{0}H_{1}H_{2}H_{3}}-iaH_{2}H_{3}}
{d\sqrt{H^{0}H_{1}H_{2}H_{3}}-icH_{2}H_{3}}=\frac{a(-i\sqrt{\frac{H_{2}H_{3}}{H^{0}H_{1}}})+b}
{c(-i\sqrt{\frac{H_{2}H_{3}}{H^{0}H_{1}}})+d}=\frac{az^{1}+b}{cz^{1}+d}\
,
\end{equation}
where we have used the fact $ad-bc=1$ repeatedly. This is in
complete agreement with (\ref{e9}). A similar check confirms that
$\tilde{z}^{2}=z^2$ and $\tilde{z}^{3}=z^3$, as required. Hence, we
can conclude that (\ref{e7}) gives the complete solutions to the
equations of motion for stationary, spherically-symmetric non-BPS
attractors. There is one small caveat, what we have actually done is
produce a general solution from a restricted subset of $D2$-$D6$
attractors with $h_0=h^1=h^2=h^3=0$; while we believe that this
simplification will not effect out result (it amounts to adding a
constant electric/magnetic potential, something which clearly does
not affect the vector equations of motion), we have been unable to
check the equations of motion for the unconstrained $D2$-$D6$ case.

\section{Split Non-BPS Attractors}
So far we've managed to establish that many of the properties of STU
non-BPS attractors are analogous to those of their BPS counterparts.
In particular, along with the basic attractor behavior established
in \cite{Goldstein:2005hq}, we have also demonstrated explicitly (at
least for the STU model) that complete expressions for the moduli
can be formed from the attractor values, by replacing charges with
harmonic functions. Given this success, it seems opportune to push a
little further.

It has been shown in \cite{Behrndt:1997ny}-\cite{Sabra:1997kq} and
\cite{Bates:2003vx}-\cite{Denef:2000nb} that static,
spherically-symmetric black holes are not the only BPS attractors
one can find. Rather the attractor mechanism can be expanded to
stationary spacetimes (dropping the requirement of spherical
symmetry) with the metric:
\begin{equation}
ds^2=-e^{2U(\mathbf{x})}\left(dt+\omega_idx^i\right)^2+e^{-2U(\mathbf{x})}dx^idx^i
\end{equation}
The most general situation described by this geometry is one
consisting of multi-centered, charged black holes. Each black hole
sits at a point $\vec{x}_a$ and carries a set of charges $\Gamma_a$.
The vector fields are then fully described by a set of harmonic
functions (a fact that doesn't rely on the BPS properties of the
black holes, or the number of centers):
\begin{equation}
H(\vec{x})= (H^{\Lambda}, H_{\Lambda}) = \mathbf{h}+
\sum_{s=1}^{n}\frac{\mathbf{\Gamma}_s}{|\vec{x}-\vec{x}_s|}\ .
\end{equation}
Bates and Denef have demonstrated (in \cite{Bates:2003vx}) that for
the BPS case these harmonic functions can be used to construct exact
solutions in an analogous fashion to the single center black holes.
$\omega_i$ is defined through the harmonic functions:
\begin{equation}
\omega_i=\left<\mathbf{H},d\mathbf{H}\right>\ .
\end{equation}
As before there is an integrability condition:
\begin{equation}
\sum_{i=1}^{n}\frac{\left<\mathbf{\Gamma_j},\mathbf{\Gamma}_i\right>}{\left|\vec{x}_j-\vec{x}_i\right|}+
\left<\mathbf{\Gamma}_j,\mathbf{h}\right>=0
\end{equation}
The above fixes the distance $\left|\vec{x}_j-\vec{x}_i\right|$
unless the charges are mutually local, with
$\left<\mathbf{\Gamma_i},\mathbf{\Gamma_j}\right>=0$ -- this is the
case when $\omega_i=0$. We also have a similar constraint on
$\mathbf{h}$ as for the single center attractor:
$\left<\mathbf{h},\mathbf{\Gamma}\right>=0$, where
$\mathbf{\Gamma}=\sum_i\mathbf{\Gamma}_i$. In this setup the BPS
solutions are given by:
\begin{eqnarray}\label{magic3}
z^\Sigma (H(\vec{x}))&=& \frac{H^{\Sigma} +i \frac{\partial
I_1(H)}{\partial H_\Sigma}}{H^0 +i \frac{\partial
I_1(H)}{\partial H_0}}\nonumber\ ,\\
e^{-2U}&=&I_1(\mathbf{H})\ .
\end{eqnarray}
Where we define the symplectic invariant $I_1$ as before:
\begin{equation}
I_1(H) = |Z(H)|^2 + |DZ(H)|^2 = \sqrt{|\mathcal{W}(H)|}\geq0\ ,
\end{equation}

For the non-BPS STU model we thus conjecture that, again analogous
to the single center black hole, solutions are given by:
\begin{eqnarray}\label{magic4}
z^\Sigma (H(\vec{x}))&=&\frac{H^{\Sigma} -i \frac{\partial
I_1(H)}{\partial H_\Sigma}}{H^0 -i \frac{\partial
I_1(H)}{\partial H_0}}\nonumber\ ,\\
e^{-2U}&=&I_1(\mathbf{H})\
\end{eqnarray}
Of course, here $\mathcal{W}(\mathbf{H})<0$. Our principle
motivation here, once again, is the underlying duality symmetry and
the constraining restrictions it must place on the form of any
solution. In theory, we can check this solution in an entirely
similar fashion to the spherically-symmetric cases discussed above.
However, the process is somewhat more involved, and as yet it has
not proved possible to carry out the required calculations.

There are, though, good reasons to believe that multi-centered,
non-BPS, extremal black hole attractors exist; and that the behavior
of the moduli fields is described by (\ref{magic4}). As discussed in
\cite{Goldstein:2005hq} and \cite{Kallosh:2006bt} the attractiveness
of black holes is a result of their extremality. Specifically,
extremal black holes have a near horizon geometry described by a
Bertotti-Robinson ($AdS_2\times S^2$) product space with an infinite
throat. It is this infinite throat that leads to attractor behavior
and it should thus be clear that if extremal multi-centered
stationary solutions can be found, then each black hole will have
the same near-horizon behavior as it would do in isolation.
Accordingly, we should expect each horizon to be an attractor.
Furthermore, it is apparent that solutions for $z^a$ of the form
(\ref{magic4}) will reduce to the single-center ones in the
near-horizon limit -- suggesting that they may well be the
appropriate global expressions for the moduli. The missing piece in
this argument, and the reason one would want to check equations of
motion explicitly, is that we cannot be certain that
\emph{stationary}, multi-centered solutions exist for extremal,
non-BPS black holes.

Before moving on to a discussion of our results, there is one last
observation that we should make. Denef has observed in
\cite{Denef:2000ar} that for a multi-center (charges $\Gamma_a$) BPS
black hole the behavior far from all centers should be that of a
spherically symmetric solution with charge $\Gamma=\sum_a\Gamma_a$,
leading to a split attractor flow. The same should hold true for
non-BPS case, though here we could have an interesting additional
effect. Whilst BPS extremal black holes are always stable, non-BPS
ones are not -- so although systems with $\Gamma_1=(q_0,0,0,p^a)$
and those with $\Gamma_2=(0,q_a,p^0,0)$ give stable attractors,
systems with $\Gamma=\Gamma_1+\Gamma_2=(q_0,q_a,p^0,p^a)$ do not.
Thus we might expect that apparently unstable ``attractors''
correspond to stable multi-centered configurations when examined at
appropriately short distances from the horizon\footnote{There is
some additional subtlety here -- since the gradient flow will have
to switch direction when the flow splits we should expect the line
of marginal stability (\cite{Denef:2000nb}, \cite{Denef:2000ar}),
where the split takes place, to have somewhat different properties
to that in the BPS case.}.

\section{Discussion}
In the preceding sections we have discussed some of issues arising
in non-BPS STU attractors. In particular we have found that the
system consisting of $p^0$ and $q_a$ charges can give rise to a
stable non-BPS attractor -- a result that was obtained using both
the attractor equation (in two different forms) and through the
minimization of an effective black hole potential. Further we have
demonstrated that one can construct exact expressions for the
moduli, the metric and the vector fields using harmonic functions
through a prescription analogous to that used for BPS black holes.

Additionally, we have argued for the existence of stationary non-BPS
attractors, with an associated split attractor flow. This allows us
to make the following conjecture regarding unstable STU non-BPS
``attractors'':
\begin{conjecture}
The apparent unstable nature of non-BPS STU ``attractors'' can be
explained through the existence of a stable multi-center solution
that is only resolved at sufficiently short distances. It is only
far away from both centers that the gradient flow of the system
appears repulsive.
\end{conjecture}
We hope to present a proof of this conjecture in future work.

Finally, we note that while everything we have discussed only
applies to STU black holes, we see no reason \emph{in principle} why
the behavior we have found and the arguments we have made for both
exact solutions and multi-center attractors will not apply to more
general extremal non-BPS black holes.

\ \leftline{\bf Acknowledgments}

This work is supported by NSF grant 0244728. N.S. and M.S. are also
supported by the U.S. Department of Energy under contract number
DE-AC02-76SF00515.

\appendix
\section{Further Details of the $D2-D6$ system}
In this appendix we discuss various details of the $D2$-$D6$ system
-- developing the solution established above by confirming it
through some alternate methods and analyzing its stability.
\subsection{Minimizing the Effective Potential}
Although we calculated the values of moduli at the black hole
horizon by using the attractor equation we can, instead, directly
extremize the effective potential; i.e. solve
$\partial_{a}V_{BH}=0$:
\begin{eqnarray}\label{b8}
\partial_{a}V_{BH}=e^{K}\Big(G^{b\bar{c}}(D_{a}D_{b}W)\bar{D}_{\bar{c}}\bar{W}+2(D_{a}W)\bar{W}\Big)\
.
\end{eqnarray}
Since the superpotential only includes odd powers of the coordinates
for the system of $D2$-$D6$ brane, we assume that the ansatz of the
minimization of the black hole potential equation is purely
imaginary $\bar{z}^{a}=-z^{a}$ (see \cite{Goldstein:2005hq}). Now,
we want to compute the ingredients of (\ref{b8}). The first
covariant derivative of the superpotential is:
\begin{equation}\label{b81}
D_{a}W=(\partial_{a}+\partial_{a}K)W=-\frac{W(z^{1},\cdots,\bar{z}^{a},\cdots,z^{3})}{z^{a}-\bar{z}^{a}}\
.
\end{equation}
Note that in the above expression, $\bar{z}^{a}$ is substituted in
the argument of the superpotential for $z^{a}$. With the ansatz
$\bar{z}^{a}=-z^{a}$, (\ref{b81}) then gives:
\begin{eqnarray}\label{b9}
D_{1}W&=&-\frac{W(\bar{z}^{1},z^{2},z^{3})}{z^{1}-\bar{z}^{1}}=\frac{W-2(q_{2}z^{2}+q_{3}z^{3})}{2z^{1}}\ ,\\
D_{2}W&=&-\frac{W(z^{1},\bar{z}^{2},z^{3})}{z^{2}-\bar{z}^{2}}=\frac{W-2(q_{1}z^{1}+q_{3}z^{3})}{2z^{2}}\ ,\\
D_{3}W&=&-\frac{W(z^{1},z^{2},\bar{z}^{3})}{z^{3}-\bar{z}^{3}}=\frac{W-2(q_{1}z^{1}+q_{2}z^{2})}{2z^{3}}\
.
\end{eqnarray}
When we drop the arguments of the superpotential they should be
understood to be $z^1$, $z^2$ and $z^3$. Before we consider the
non-SUSY solutions, we find the supersymmetric ones, $D_{a}W=0$:
\begin{equation}\label{b10}
z^1=i\sqrt{-\frac{q_{2}q_{3}}{p^{0}q_{1}}}\
,z^2=i\sqrt{-\frac{q_{1}q_{3}}{p^{0}q_{2}}}\
,z^3=i\sqrt{-\frac{q_{1}q_{2}}{p^{0}q_{3}}}\ ,
\end{equation}
and also
\begin{equation}\label{b101}
z^1=-i\sqrt{-\frac{q_{2}q_{3}}{p^{0}q_{1}}}\
,z^2=-i\sqrt{-\frac{q_{1}q_{3}}{p^{0}q_{2}}}\
,z^3=-i\sqrt{-\frac{q_{1}q_{2}}{p^{0}q_{3}}}\ .
\end{equation}
As briefly discussed above, we should examine these solutions to see
whether they respect the positivity of the K\"{a}hler potential -- a
necessary condition to ensure positive kinetic terms in the action.
A simple check shows that the first SUSY solution (\ref{b10})
violates the positivity of the K\"{a}hler potential and therefore,
it is ruled out. Thus only the latter (\ref{b101}) is acceptable. We
should also notice that these solutions will only exist when
$p^{0}q_{1}q_{2}q_{3}<0$.

The next object to calculate is the second covariant derivative of
$W$:
\begin{eqnarray}\label{b11}
D_{a}D_{b}W=(\partial_{a}+\partial_{a}K)D_{b}W-\Gamma^{c}_{ab}D_{c}W\
.
\end{eqnarray}
For our particular STU model:
\begin{equation}\label{b121}
D_{a}D_{b}W=(1-\delta_{ab})\frac{W(z^{1},\cdots,\bar{z}^{a},\bar{z}^{b},\cdots)}
{(z^{a}-\bar{z}^{a})(z^{b}-\bar{z}^{b})}\ .
\end{equation}
Notice that $\bar{z}^{a}$ and $\bar{z}^{b}$ substitute $z^{a}$ and
$z^{b}$ in the argument of $W$. For the $z^a=-\bar{z}^a$ ansatz of
our $D2$-$D6$ brane system, the above reduces to:
\begin{equation}\label{b12}
D_{a}D_{b}W=(1-\delta_{ab})\frac{W-2(q_{a}z^{a}+q_{b}z^{b})}{4z^{a}z^{b}}\
\ (\mbox{no sum on}\ a\ \mbox{and}\ b)\ .
\end{equation}
Now, we are ready to form the equation ofan explicit expression for
the minimization of the potential $\partial_{a}V_{BH}=0$. Using
(\ref{b3}), (\ref{b9}), and (\ref{b12}), we find the following set
of equations:
\begin{eqnarray}\label{b13}
&
\Big(W-2(q_{1}z^{1}+q_{2}z^{2})\Big)\Big(W-2(q_{1}z^{1}+q_{3}z^{3})\Big)
+W\Big(W-2(q_{2}z^{2}+q_{3}z^{3})\Big)=0\ ,\\
&
\Big(W-2(q_{1}z^{1}+q_{2}z^{2})\Big)\Big(W-2(q_{2}z^{2}+q_{3}z^{3})\Big)
+W\Big(W-2(q_{1}z^{1}+q_{3}z^{3})\Big)=0\ ,\\
&
\Big(W-2(q_{1}z^{1}+q_{3}z^{3})\Big)\Big(W-2(q_{2}z^{2}+q_{3}z^{3})\Big)
+W\Big(W-2(q_{1}z^{1}+q_{2}z^{2})\Big)=0\ .
\end{eqnarray}
It is clear that (\ref{b101}) satisfy these equations. Now, we want
to find the non-SUSY solutions ($D_{a}W\neq 0$). If we divide the
above expressions by one another, we obtain:
\begin{eqnarray}\label{b14}
\Big(W-2(q_{1}z^{1}+q_{2}z^{2})\Big)^{2}=\Big(W-2(q_{2}z^{2}+q_{3}z^{3})\Big)^{2}
=\Big(W-2(q_{1}z^{1}+q_{3}z^{3})\Big)^{2}\ .
\end{eqnarray}
The complete solutions to these relations are:
\begin{eqnarray}\label{b171}
z^{1}=\pm i\sqrt{\frac{q_{2}q_{3}}{p^{0}q_{1}}}\ ,\ z^{2}=\pm
i\sqrt{\frac{q_{1}q_{3}}{p^{0}q_{2}}}\ ,\ z^{3}=\pm
i\sqrt{\frac{q_{1}q_{2}}{p^{0}q_{3}}}\ ,
\end{eqnarray}
If we consider all these possibilities for (\ref{b14}), then we find
eight sets, which can be categorized into two groups:
\begin{eqnarray}\label{b191}
&A:\ \{(+,+,+),(+,-,-),(-,+,-),(-,-,+)\}\ ,\\
&B:\ \{(-,-,-),(-,+,+),(+,-,+),(+,+,-)\}\ .
\end{eqnarray}
In above, each parenthesis shows the sign of $(z^{1},z^{2},z^{3})$
respectively. It turns out that the elements of group $A$ violate
the positivity of $e^K$ and therefore this group is unacceptable.
However, no such problem exists for group $B$, and therefore its
elements provide acceptable non-SUSY solutions. Furthermore, the
last three elements of group $B$ are physically equivalent solutions
in which the sign of two fields are the same, but opposite to the
third. It's worth noting that that the attractor equation gave us
only the $(-,-,-)$ solution; however, this point has little effect
on the rest of out argument.

Finally, we note that for all non-SUSY solutions, we must have
$p^{0}q_{1}q_{2}q_{3}>0$. This in turn means that we cannot
simultaneously have both SUSY and non-SUSY solutions.

\subsection{An Alternate Attractor Equation}
In this section, we find the values of the moduli at the horizon of
the black hole by solving the attractor equation in the form given
in \cite{Kallosh:2006bt}:
\begin{eqnarray}\label{d1}
\Gamma=2\mbox{Im}\Big[Z\bar{\Pi}-\frac{(\bar{D}_{\bar{a}}\bar{D}_{\bar{b}}\bar{Z})
G^{\bar{a}c}G^{\bar{b}d}D_{c}Z}{2Z}D_{d}\Pi\Big]\ .
\end{eqnarray}
As before $\Gamma$ is the set of magnetic and electric charges
$\Gamma=\left(p^{\Lambda},q_{\Lambda}\right)$ and $\Pi$ is the
covariantly holomorphic period vector:
\begin{eqnarray}\label{d2}
\Pi=e^{K/2}\left(
             \begin{array}{c}
               1 \\
               z^{a} \\
               F_{0} \\
               F_{a} \\
             \end{array}
           \right)\ .
\end{eqnarray}
We have already calculated most of these ingredients, with our only
remaining task to calculate the covariant derivative of $\Pi$:
\begin{eqnarray}\label{d3}
D_{d}\Pi=e^{K/2}\left(
                  \begin{array}{c}
                    \partial_{d}K \\
                    \delta^{a}_{d}+(\partial_{d}K)z^{a} \\
                    \partial_{d}F_{0}+(\partial_{d}K)F_{0} \\
                    \partial_{d}F_{a}+(\partial_{d}K)F_{a} \\
                  \end{array}
                \right)\ .
\end{eqnarray}
It is straightforward to calculate the two terms of the attractor
equation for our ansatz $\bar{z}^{a}=-z^{a}$. These are:
\begin{eqnarray}\label{d4}
2\mbox{Im}(Z\bar{\Pi})=\frac{W}{4z^{1}z^{2}z^{3}} \left(
  \begin{array}{c}
    1 \\
    0 \\
    0 \\
    0 \\
    0 \\
    z^{2}z^{3} \\
    z^{1}z^{3} \\
    z^{1}z^{2} \\
  \end{array}
\right)\ , \
2\mbox{Im}\Big[\frac{(\bar{D}_{\bar{a}}\bar{D}_{\bar{b}}\bar{Z})
G^{\bar{a}c}G^{\bar{b}d}D_{c}Z}{2Z}D_{d}\Pi\Big]=-\frac{1}{4z^{1}z^{2}z^{3}}\frac{1}{W}
\left(
  \begin{array}{c}
    Y_{0} \\
    0 \\
    0 \\
    0 \\
    0 \\
    Y_{1}z^{2}z^{3} \\
    Y_{2}z^{1}z^{3} \\
    Y_{3}z^{1}z^{2} \\
  \end{array}
\right)\ .\nonumber\\
\
\end{eqnarray}
We have defined $Y_{0}$, $Y_{1}$, $Y_{2}$, and $Y_{3}$ to be:
\begin{eqnarray}\label{d6}
Y_{0}&\equiv& W(\bar{z}^{1},z^{2},z^{3})W(z^{1},\bar{z}^{2},z^{3})
+W(\bar{z}^{1},z^{2},z^{3})W(z^{1},z^{2},\bar{z}^{3})
+W(z^{1},\bar{z}^{2},z^{3})W(z^{1},z^{2},\bar{z}^{3})\nonumber\\
Y_{1}&\equiv& W(\bar{z}^{1},z^{2},z^{3})W(z^{1},\bar{z}^{2},z^{3})
+W(\bar{z}^{1},z^{2},z^{3})W(z^{1},z^{2},\bar{z}^{3})
-W(z^{1},\bar{z}^{2},z^{3})W(z^{1},z^{2},\bar{z}^{3})\nonumber\\
Y_{2}&\equiv& W(\bar{z}^{1},z^{2},z^{3})W(z^{1},\bar{z}^{2},z^{3})
+W(z^{1},\bar{z}^{2},z^{3})W(z^{1},z^{2},\bar{z}^{3})
-W(\bar{z}^{1},z^{2},z^{3})W(z^{1},z^{2},\bar{z}^{3})\nonumber\\
Y_{3}&\equiv& W(\bar{z}^{1},z^{2},z^{3})W(z^{1},z^{2},\bar{z}^{3})
+W(z^{1},\bar{z}^{2},z^{3})W(z^{1},z^{2},\bar{z}^{3})-
W(\bar{z}^{1},z^{2},z^{3})W(z^{1},\bar{z}^{2},z^{3})\ .\nonumber\\
&&
\end{eqnarray}
Forming the attractor equation (\ref{d1}), we find the following:
\begin{eqnarray}\label{d7}
&&W^{2}-Y_{0}=4p^{0}Wz^{1}z^{2}z^{3}\\
&&W^{2}+Y_{a}=4q_{a}Wz^{a}\ (\mbox{no sum on\ }a)\ .
\end{eqnarray}
As is clear, we have four equations, but only three of them are
independent. This can be easily seen if we add the four to one
another and use $Y_{1}+Y_{2}+Y_{3}=Y_{0}$. To solve these equations
we first note that
$Y_{1}-4q_{1}Wz^{1}=Y_{2}-4q_{2}Wz^{2}=Y_{3}-4q_{3}Wz^{3}$ and
observe this leads to:
\begin{eqnarray}\label{d81}
q_{1}z^{1}=\pm q_{2}z^{2}=\pm q_{3}z^{3}\ .
\end{eqnarray}
If we take $q_{1}z^{1}=q_{2}z^{2}=q_{3}z^{3}$ we find:
\begin{eqnarray}\label{d8}
\left(z^{1}-\frac{p^{0}q_{1}}{q_{2}q_{3}}(z^{1})^{3}\right)
\left(z^{1}+\frac{p^{0}q_{1}}{q_{2}q_{3}}(z^{1})^{3}\right)=0\ .
\end{eqnarray}
We immediately realize that the first parenthesis produces the
supersymmetric solution:
\begin{eqnarray}\label{d9}
z^{1}=\pm i\sqrt{-\frac{q_{2}q_{3}}{p^{0}q_{1}}}\ ,
\end{eqnarray}
The second parenthesis of (\ref{d8}) then gives us the
non-supersymmetric attractor:
\begin{eqnarray}\label{d10}
z^{1}=\pm i\sqrt{\frac{q_{2}q_{3}}{p^{0}q_{1}}}\ .
\end{eqnarray}
As we see, different choices of (\ref{d81}) produce all the
different SUSY and non-SUSY solutions that we found in the previous
section. Recalling that the positivity of $e^K$ requires we choose a
specific set of signs for the above solutions leads to (\ref{b101})
for the supersymmetric solution and elements of group $B$ (see
\ref{b191}) for the non-supersymmetric ones.

\subsection{Stability}
In the previous section, we found both SUSY and non-SUSY attractors.
The former are always stable, since the second derivative of the
black hole potential is proportional to the K\"{a}hler metric.
However, for non-SUSY solutions, we need to check stability
explicitly. In order to do this, we need to find the full mass
matrix.

First, note that at the extremum of the potential we have
$\partial_{a}\partial_{b}V_{BH}=D_{a}D_{b}V_{BH}$; therefore, we can
use the covariant rather than ordinary derivative. Then the
holomorphic-holomorphic and holomorphic-antiholomorphic parts of the
mass matrix are given by
\begin{eqnarray}\label{c1}
D_{a}D_{b}V_{BH}&=&e^{K}\Big(G^{c\bar{d}}(D_{a}D_{b}D_{c}W)\bar{D}_{\bar{d}}\bar{W}
+3(D_{a}D_{b}W)\bar{W}\Big)\ ,\\
\bar{D}_{\bar{a}}D_{b}V_{BH}&=&e^{K}\Big(-{\mathcal{R}}^{d}_{b\bar{a}c}G^{c\bar{e}}(D_{d}W)\bar{D}_{\bar{e}}\bar{W}
+G^{c\bar{d}}(D_{b}D_{c}W)
(\bar{D}_{\bar{a}}\bar{D}_{\bar{d}}\bar{W})\nonumber\\
&&+G_{\bar{a}b}G^{c\bar{d}}(D_{c}W)\bar{D}_{\bar{d}}\bar{W}
+3(\bar{D}_{\bar{a}}\bar{W})D_{b}W+2G_{\bar{a}b}W\bar{W}\Big)\ ,
\end{eqnarray}
where ${\mathcal{R}}^{d}_{b\bar{a}c}$ is the Riemann curvature
tensor of the moduli space. We begin with the
holomorphic-holomorphic piece. This requires us to compute the third
covariant derivative of the superpotential:
\begin{eqnarray}\label{c2}
D_{a}D_{b}D_{c}W=-|\epsilon_{abc}|\frac{W(\bar{z}^{1},\bar{z}^{2},\bar{z}^{3})}
{(z^{a}-\bar{z}^{a})(z^{b}-\bar{z}^{b})(z^{c}-\bar{z}^{c})}\ , \ \
(\mbox{no sum on\ } a,b, \mbox{and\ } c)\ .
\end{eqnarray}
Using this result and defining $B_{ab}\equiv
\left.D_{a}D_{b}V_{BH}\right|_{horizon}$ we have:
\begin{eqnarray}\label{c3}
B=\frac{1}{2}p^{0}\sqrt{p^{0}q_{1}q_{2}q_{3}} \left(
  \begin{array}{ccc}
    0 & \frac{1}{q_{3}} & \frac{1}{q_{2}} \\
    \frac{1}{q_{3}} & 0 & \frac{1}{q_{1}} \\
    \frac{1}{q_{2}} & \frac{1}{q_{1}} & 0 \\
  \end{array}
\right)\ .
\end{eqnarray}
To find the antiholomorphic-holomorphic piece of the mass matrix, we
need the Riemann curvature of the moduli space:
\begin{eqnarray}\label{c4}
{\mathcal{R}}^{a}_{a\bar{a}a}=-\frac{2}{(z^{a}-\bar{z}^{a})^{2}}\ ,\
\ \ (\mbox{no sum on\ }a)\ .
\end{eqnarray}
All other components of the Riemann curvature tensor vanish. Taking
this result we can crawl through a long computation to obtain
$A_{ab}=\bar{D}_{\bar{a}}D_{b}V_{BH}\Big|_{horizon}$ as:
\begin{eqnarray}\label{c5}
A=\frac{1}{2}p^{0}\sqrt{p^{0}q_{1}q_{2}q_{3}} \left(
  \begin{array}{ccc}
    \frac{2q_{1}}{q_{2}q_{3}} & \frac{1}{q_{3}} & \frac{1}{q_{2}} \\
    \frac{1}{q_{3}} & \frac{2q_{2}}{q_{1}q_{3}} & \frac{1}{q_{1}} \\
    \frac{1}{q_{2}} & \frac{1}{q_{1}} & \frac{2q_{3}}{q_{1}q_{2}} \\
  \end{array}
\right)\ .
\end{eqnarray}
Hence, the full mass matrix of the black hole potential is given by:
\begin{eqnarray}\label{c6}
M=\left(
  \begin{array}{cc}
    A & \bar{B} \\
    B & \bar{A} \\
  \end{array}
\right)=\frac{1}{2}p^{0}\sqrt{p^{0}q_{1}q_{2}q_{3}} \left(
  \begin{array}{cccccc}
    \frac{2q_{1}}{q_{2}q_{3}} & \frac{1}{q_{3}} & \frac{1}{q_{2}} & 0 & \frac{1}{q_{3}} & \frac{1}{q_{2}} \\
    \frac{1}{q_{3}} & \frac{2q_{2}}{q_{1}q_{3}} & \frac{1}{q_{1}} & \frac{1}{q_{3}} & 0 & \frac{1}{q_{1}} \\
    \frac{1}{q_{2}} & \frac{1}{q_{1}} & \frac{2q_{3}}{q_{1}q_{2}} & \frac{1}{q_{2}} & \frac{1}{q_{1}} & 0 \\
    0 & \frac{1}{q_{3}} & \frac{1}{q_{2}} & \frac{2q_{1}}{q_{2}q_{3}} & \frac{1}{q_{3}} & \frac{1}{q_{2}} \\
    \frac{1}{q_{3}} & 0 & \frac{1}{q_{1}} & \frac{1}{q_{3}} & \frac{2q_{2}}{q_{1}q_{3}} & \frac{1}{q_{1}} \\
    \frac{1}{q_{2}} & \frac{1}{q_{1}} & 0 & \frac{1}{q_{2}} & \frac{1}{q_{1}} & \frac{2q_{3}}{q_{1}q_{2}} \\
  \end{array}
\right)\ .
\end{eqnarray}
Note that this will give rise to a term in the effective potential
which takes the following form:
\begin{eqnarray}\label{massterm}
z^{\dag}Mz\qquad\textrm{where,}\quad
z=\left(z^1,z^2,z^3,\bar{z}^1,\bar{z}^2,\bar{z}^3\right)\ .
\end{eqnarray}
To explore the stability of the non-SUSY solutions we need to
diagonalize the mass matrix. This is straightforward and the
eigenvalues are given by:
\begin{eqnarray}\label{c7}
\left\{0,0,\sqrt{p^{0}q_{1}q_{2}q_{3}}\frac{p^{0}q_{1}}{q_{2}q_{3}},\sqrt{p^{0}q_{1}q_{2}q_{3}}
\frac{p^{0}q_{2}}{q_{1}q_{3}},\sqrt{p^{0}q_{1}q_{2}q_{3}}\frac{p^{0}q_{3}}{q_{1}q_{2}},\sqrt{p^{0}q_{1}q_{2}q_{3}}
\Big(\frac{p^{0}q_{1}}{q_{2}q_{3}}+\frac{p^{0}q_{2}}{q_{1}q_{3}}+\frac{p^{0}q_{3}}{q_{1}q_{2}}\Big)\right\}\
.
\end{eqnarray}
There are two flat directions, with the rest of the eigenvalues
positive. For the former we find that the associated eigenvectors
are:
\begin{eqnarray}\label{nav2}
\left(-\frac{q_3}{q_1},0,1,-\frac{q_3}{q_1},0,1\right)\qquad\textrm{and}\qquad
\left(-\frac{q_2}{q_1},1,0,-\frac{q_2}{q_1},1,0\right)\ .
\end{eqnarray}
Both of these (as is clear from (\ref{massterm})) are in a real
direction -- moving along potential in the direction of the
eigenvectors sends $z^a\rightarrow z^a+\delta x^a$, where $\delta
x^a$ is real.

In order to explore the stability of our extrema with respect to the
flat directions, we need to examine higher order derivatives. Recall
that we initially proceeded with covariant derivatives instead of
ordinary derivatives (since $DDV_{BH}=d^{2}V_{BH}$ at extremal
points of $V_{BH}$). But this is no longer true at higher orders, so
one needs to revert to the flat derivative. Unfortunately, this
becomes rather unpleasant, as we forced to break the covariant form,
hence we choose to proceed in a calculationally more tractable
fashion: We expand the black hole potential around the
non-supersymmetric solutions and then concentrate on the behavior of
cubic and quartic terms of the expansion. If the coefficients of the
cubic terms are non-vanishing, then we can conclude that the
non-supersymmetric solution is unstable. However, if the
coefficients vanish, then we need to consider the quartic terms. If
these terms are positive for flat directions, independent of the
values of the parameter of the expansion, then the
non-supersymmetric solution is indeed stable.

First, using (\ref{b3}) and (\ref{b9}), we notice that we can
rewrite the black hole potential in the following way
\begin{equation}\label{c10}
V_{BH}=e^{K}\Big(|W(\bar{z}^{1},z^{2},z^{3})|^{2}+|W(z^{1},\bar{z}^{2},z^{3})|^{2}
+|W(z^{1},z^{2},\bar{z}^{3})|^{2}+|W(z^{1},z^{2},z^{3})|^{2}\Big)\ .
\end{equation}
Now, we want to expand the above potential around the
non-supersymmetric solution (\ref{b171}) as
\begin{eqnarray}\label{c11}
z^{1}=-i\sqrt{\frac{q_{2}q_{3}}{p^{0}q_{1}}}+\epsilon^{1}\ ,\ \
z^{2}=-i\sqrt{\frac{q_{1}q_{3}}{p^{0}q_{2}}}+\epsilon^{2}\ ,\ \
z^{3}=-i\sqrt{\frac{q_{1}q_{2}}{p^{0}q_{3}}}+\epsilon^{3}\ ,
\end{eqnarray}
where $\epsilon^{a}$ is a real parameter of the expansion. Why only
real? Well we are, of course, only concerned with the quartic
expansion in the flat directions, and we have already established
that, for our first non-supersymmetric solution, these are real.
Accordingly $V_{BH}$ has the following expansion
\begin{eqnarray}\label{c12}
V_{BH}(z^{a})=V_{BH}(z_{0}^{a})+{\mathcal{O}}((\epsilon^{a})^{2})+{\mathcal{O}}((\epsilon^{a})^{3})
+{\mathcal{O}}((\epsilon^{a})^{4})+\cdots\ ,
\end{eqnarray}
where $z_{0}^{a}$ is the value of moduli for non-supersymmetric
solution. Note that the flat directions in (\ref{nav2}) correspond
to the plane $q_1 \epsilon^1+q_2 \epsilon^2 +q_3 \epsilon^3=0$ in
notation of (\ref{c11}).

The information in ${\mathcal{O}}((\epsilon^{a})^{2})$ is encoded in
the mass matrix which has already been calculated and we need to
know about higher order terms along flat directions. Explicit
computation shows that the cubic terms vanish\footnote{\ This result
is no longer true in the presence of D4 branes. In fact, the details
of the calculations show that if all existing terms in the
superpotential either have even or odd powers of the moduli
coordinates, then the cubic term ${\mathcal{O}}((\epsilon^{a})^{3})$
(which is the leading term in perturbation) vanishes. But in general
case when the superpotential has a mixture of even and odd powers,
then the cubic term does not vanish and therefore, the extremum is
an inflection point rather than a minimum.}
${\mathcal{O}}((\epsilon^{a})^{3})=0$. Thus we must examine the
quartic terms and we find:
\begin{eqnarray}\label{c13}
{\mathcal{O}}((\epsilon^{a})^{4})&=&4e^{K(z_{0}^{a})}p^{0}\Big(\frac{q_{1}q_{2}}{q_{3}}
(\epsilon^{1}\epsilon^{2})^{2}+\frac{q_{2}q_{3}}{q_{1}}(\epsilon^{2}\epsilon^{3})^{2}
+\frac{q_{1}q_{3}}{q_{2}}(\epsilon^{1}\epsilon^{3})^{2}\nonumber\\
&&+q_{1}(\epsilon^{1})^{2}\epsilon^{2}\epsilon^{3}
+q_{2}\epsilon^{1}(\epsilon^{2})^{2}\epsilon^{3}+q_{3}\epsilon^{1}\epsilon^{2}(\epsilon^{3})^{2}\Big)\nonumber\\
&=&2e^{K(z_{0}^{a})}\frac{(p^0)^2}{p^0q_1q_2q_3}\left[a_1^2+
a_2^2+a_3^2 +(a_1+ a_2 + a_3)^2\right] \ .
\end{eqnarray}
Here $a_1=q_2\epsilon^2q_3\epsilon^3$,
$a_2=q_1\epsilon^1q_3\epsilon^3$ and
$a_3=q_1\epsilon^1q_2\epsilon^2$. As we see, no matter how the
parameters of expansion change, ${\mathcal{O}}((\epsilon^{a})^{4})$
is manifestly positive.

This implies that the non-SUSY solution (\ref{b171}) is indeed
stable. Finally, we mention that it can be (and has been) checked
that the other non-SUSY solution $(-, +, +)$ also gives a positive
mass matrix.

\end{document}